\documentclass{appolb}
\usepackage{graphicx}

\begin{document}
\title{Exploring double-parton scattering effects for jets \\
 with large rapidity separation \\ 
and in four-jet production at the LHC%
\thanks{Presented at the 16th conference on Elastic and Diffractive scattering, EDS Blois 2015 }%
}
\author{Antoni Szczurek
\address{H.Niewodnicza\'nski Institute of Nuclear Physics, Polish Academy of Sciences, Radzikowskiego 152, 31-342 Krak\'ow, Poland \and University of Rzesz\'ow, Rejtana 16, 35-959 Rzesz\'ow, Poland}
}
\maketitle
\begin{abstract}
We present an estimation of the contribution of double parton scattering
(DPS) for jets widely separated in rapidity and for four-jet sample.
In the case of four-jet production we calculate cross section for 
both single-parton scattering (SPS) using the code ALPGEN as well as 
for DPS in LO collinear approach.
The DPS contribution is calculated within the so-called factorized
Ansatz and each step of DPS is calculated in the LO collinear approximation. 
We show that the relative (with respect to SPS dijets and to the
BFKL Mueller-Navelet (MN) jets) contribution of DPS is growing at 
large rapidity distance between jets. 
The calculated differential cross sections as a function of rapidity
distance between the most remote in rapidity jets are compared with 
recent results of LL and NLL BFKL calculations for the Mueller-Navelet 
jet production at $\sqrt{s} = 7$ TeV.  
Our results for four-jet sample are compared
with experimental data obtained recently by the CMS collaboration
and a rather good agreement is achieved.
We propose to impose different cuts
in order to enhance the relative DPS contribution.
The relative DPS contribution increases when decreasing the lower
cut on the jet transverse momenta as well as when a low lower cut
on the rapidity distance between the most remote jets is imposed.
\end{abstract}
\PACS{11.80.La,13.87.Ce,14.65.Dw,14.70.Fm}

\section{Introduction}

\begin{figure}[!h]
\begin{center}
\includegraphics[width=4.0cm]{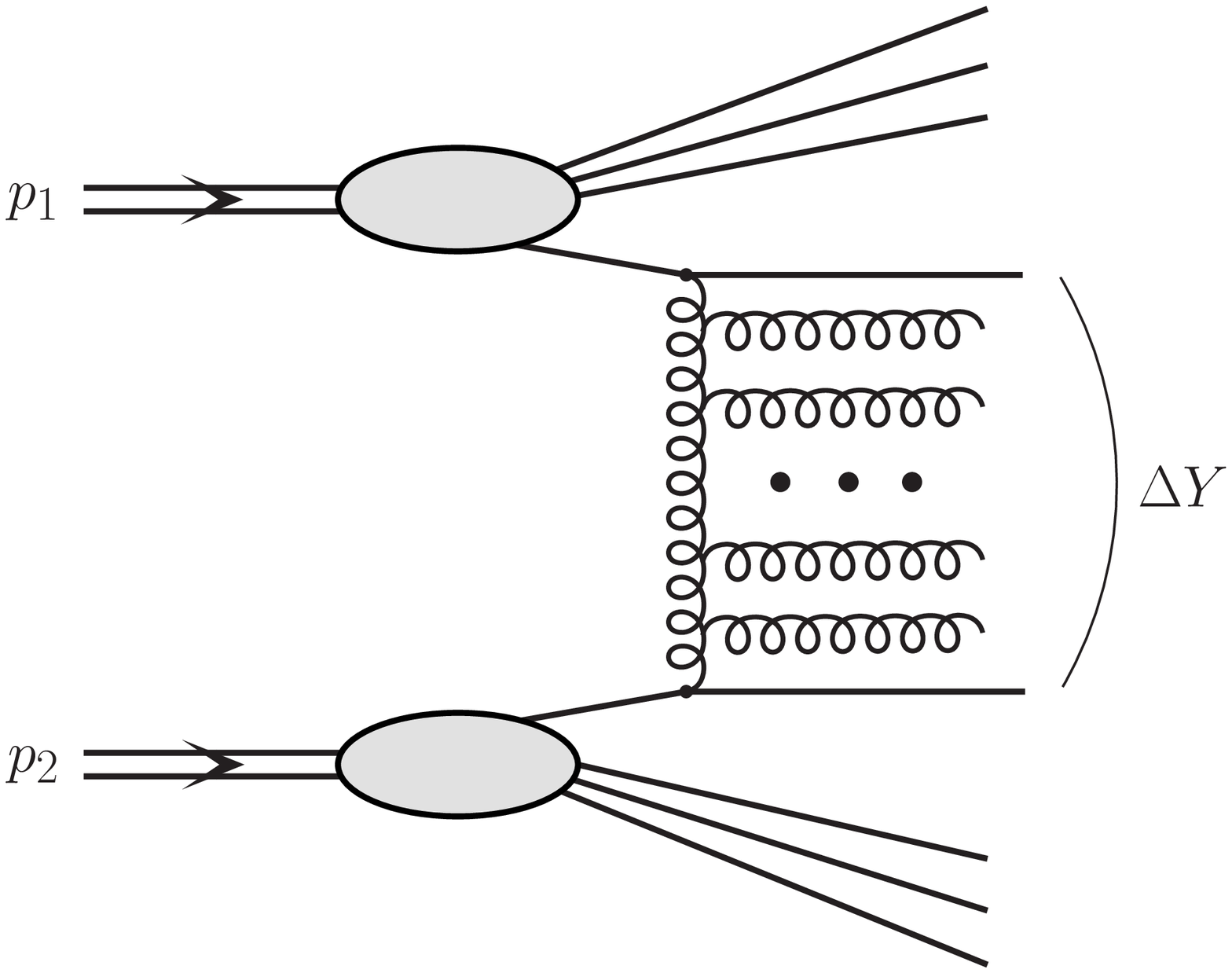}
\includegraphics[width=4.0cm]{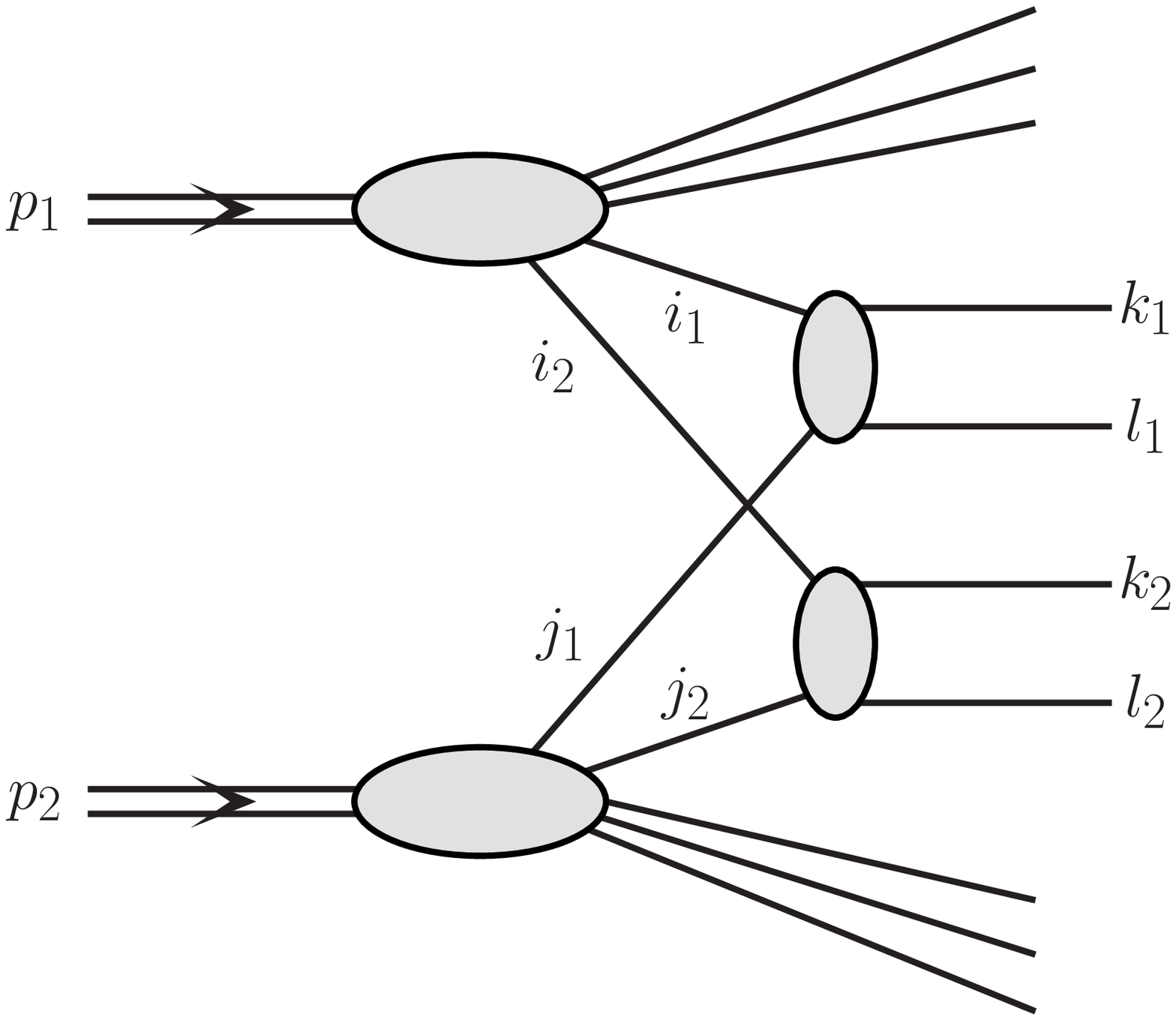}
\end{center}
\vskip -0.3cm
\caption{
\small A diagramatic representation of the Mueller-Navelet jet
production (left diagram) and of the double paron scattering mechanism
(right diagram).
}
 \label{fig:diagrams}
\end{figure}

Many years ago Mueller and Navelet predicted strong decorrelation 
in relative azimuthal angle \cite{Mueller:1986ey} of jets with large
rapidity separation due to exchange of the BFKL ladder between quarks
(see left panel of Fig.\ref{fig:diagrams}). 
Since then both leading-logarithmic
and higher-order BFKL effects
were calculated and discussed.
The effect of the NLL correction is large and leads to significant
lowering of the cross section.
The LHC opens a new possibility to study the decorrelation in azimuthal angle. 
First experimental data measured at $\sqrt{s}$ = 7 TeV are expected 
soon \cite{CMS_private}.
Also double parton scattering (DPS) can be important in this context
(for diagrammatic representation of DPS see right panel of 
Fig.\ref{fig:diagrams}).
We discussed recently how important is the contribution of DPS
in the case of the jets widely separated in rapidity \cite{MS2014}
and for four-jet sample \cite{MS2015}.

Four-jet production was already discussed in the context of
double parton scattering. Actually it was a first process where
the DPS was claimed to be observed experimentally \cite{Tevatron_4jets}.
However, in most of the past as well as current analyses the DPS
contribution to four-jet production is relatively small and single
parton scattering (SPS) driven by the 2 $\to$ 4 partonic processes
dominates. 

On the theoretical side the DPS effects in four-jet production were 
discussed in 
Refs.~\cite{Treleani_jet1,Treleani_jet2,Mangano,Domdey,Berger,Strikman}.
A first theoretical estimate of SPS four-jet production, including only
some partonic subprocesses, and its comparison to DPS contribution was
presented in Ref.~\cite{Mangano} for Tevatron.
Some new kinematical variables useful for identification of DPS
were proposed in Ref.~\cite{Berger}.
Presence of perturbative parton splitting mechanism 
was discussed in Ref.~\cite{Strikman}.

In our recent studies we have shown how big can be the contribution of
DPS for jets widely separated in rapidity \cite{MS2014}.
Understanding of this contribution is important in the context of 
searching for BFKL effects or in general QCD higher-order effects
\cite{Chatrchyan:2013qza}. 


In the present letter we wish to discuss also exclusive four-jet sample
where the situation in the context of searching for DPS is 
even better \cite{MS2015}. In Ref.\cite{MS2015} we showed how to
maximize the DPS contribution by selecting relevant kinematical cuts.
Here we shall show only some examples.

\section{DPS mechanism}

Partonic cross sections used to calculate DPS are calculated 
only in leading order.
Then the cross section for dijet production can be written as:
\begin{equation}
\frac{d \sigma(i j \to k l)}{d y_1 d y_2 d^2p_t} 
= \frac{1}{16 \pi^2 {\hat s}^2}
\sum_{i,j} x_1 f_i(x_1,\mu^2) \; x_2 f_j(x_2,\mu^2) \;
\overline{|\mathcal{M}_{i j \to k l}|^2} \;.
\label{LO_SPS}
\end{equation}
In our calculations we include all leading-order $i j \to k l$ partonic 
subprocesses.
The $K$-factor for dijet production is rather small, of the order of 
$1.1 - 1.3$. 
It was shown in Ref.\cite{MS2014} that
already the leading-order approach gives results in sufficiently reasonable 
agreement with recent ATLAS and CMS inclusive jet data.

This simplified leading-order approach can be however used easily in 
calculating DPS differential cross sections. 
The multi-dimensional differential cross section can be written as:
\begin{equation}
\frac{d \sigma^{DPS}(p p \to \textrm{4jets} \; X)}{d y_1 d y_2 d^2p_{1t}
  d y_3 d y_4 d^2p_{2t}} 
= \sum_{i_1,j_1,k_1,l_1;i_2,j_2,k_2,l_2} \; 
\frac{\mathcal{C}}{\sigma_{eff}} \;
\frac{d \sigma(i_1 j_1 \to k_1 l_1)}{d y_1 d y_2 d^2p_{1t}} \; 
\frac{d \sigma(i_2 j_2 \to k_2 l_2)}{d y_3 d y_4 d^2p_{2t}}\;, 
\label{DPS}
\end{equation}
where
$\mathcal{C} = \left\{ \begin{array}{ll}
\frac{1}{2}\;\; & \textrm{if} \;\;i_1 j_1 = i_2 j_2 \wedge k_1 l_1 = k_2 l_2\\
1\;\;           & \textrm{if} \;\;i_1 j_1 \neq i_2 j_2 \vee k_1 l_1 \neq k_2 l_2
\end{array} \right\} $ and partons 
$j,k,l,m = g, u, d, s, \bar u, \bar d, \bar s$. 
The combinatorial factors include identity of the two subprocesses.
Each step of DPS is calculated in the leading-order approach 
(see Eq.(\ref{LO_SPS})).

In the calculations we have taken in most cases $\sigma_{eff}$ = 15 mb.
Phenomenological studies of $\sigma_{eff}$ summarized e.g. in
\cite{Seymour} give a similar value.

\section{DPS and jets with large rapidity separation}

\begin{figure}[!h]
\begin{center}
\includegraphics[width=5cm]{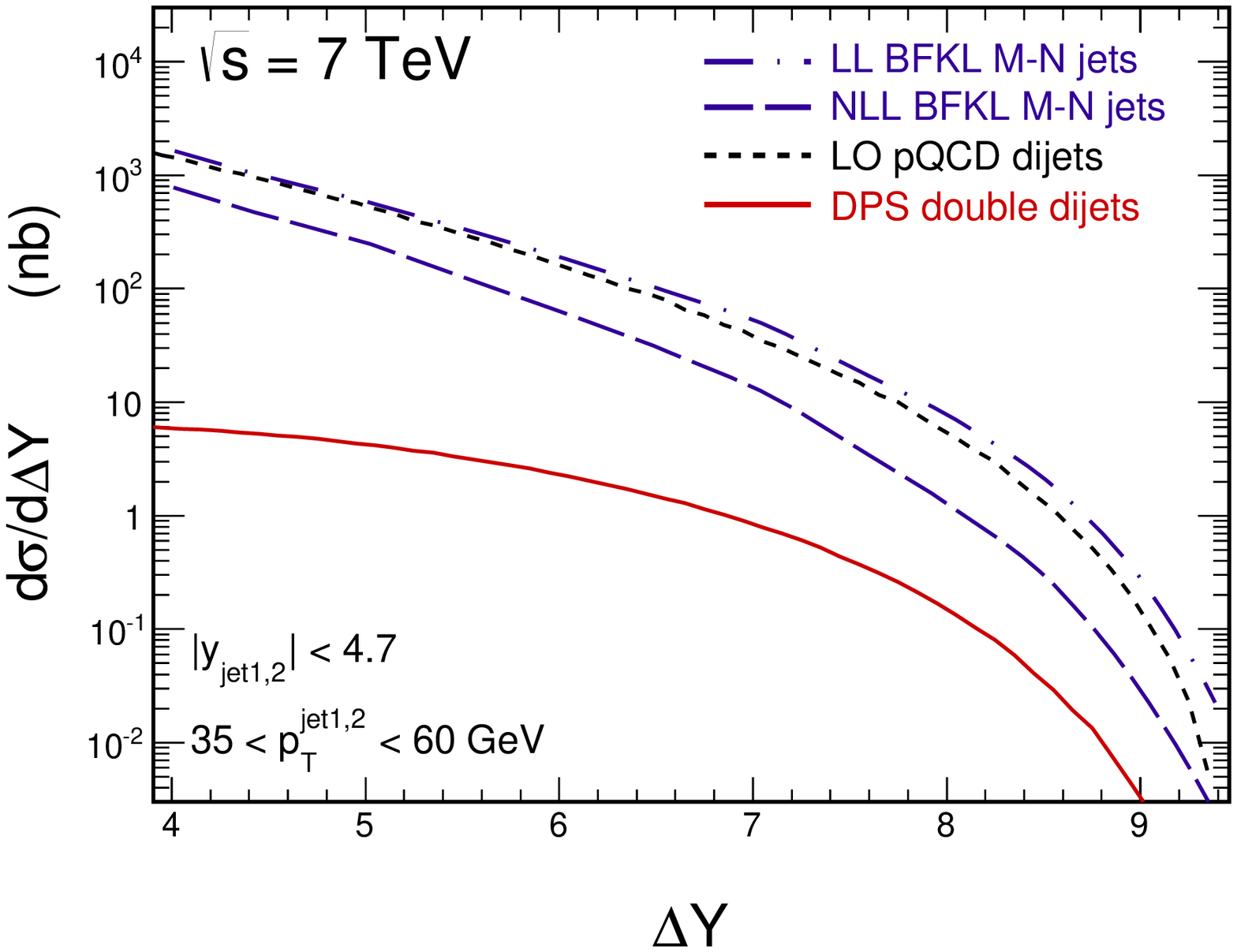}
\includegraphics[width=5cm]{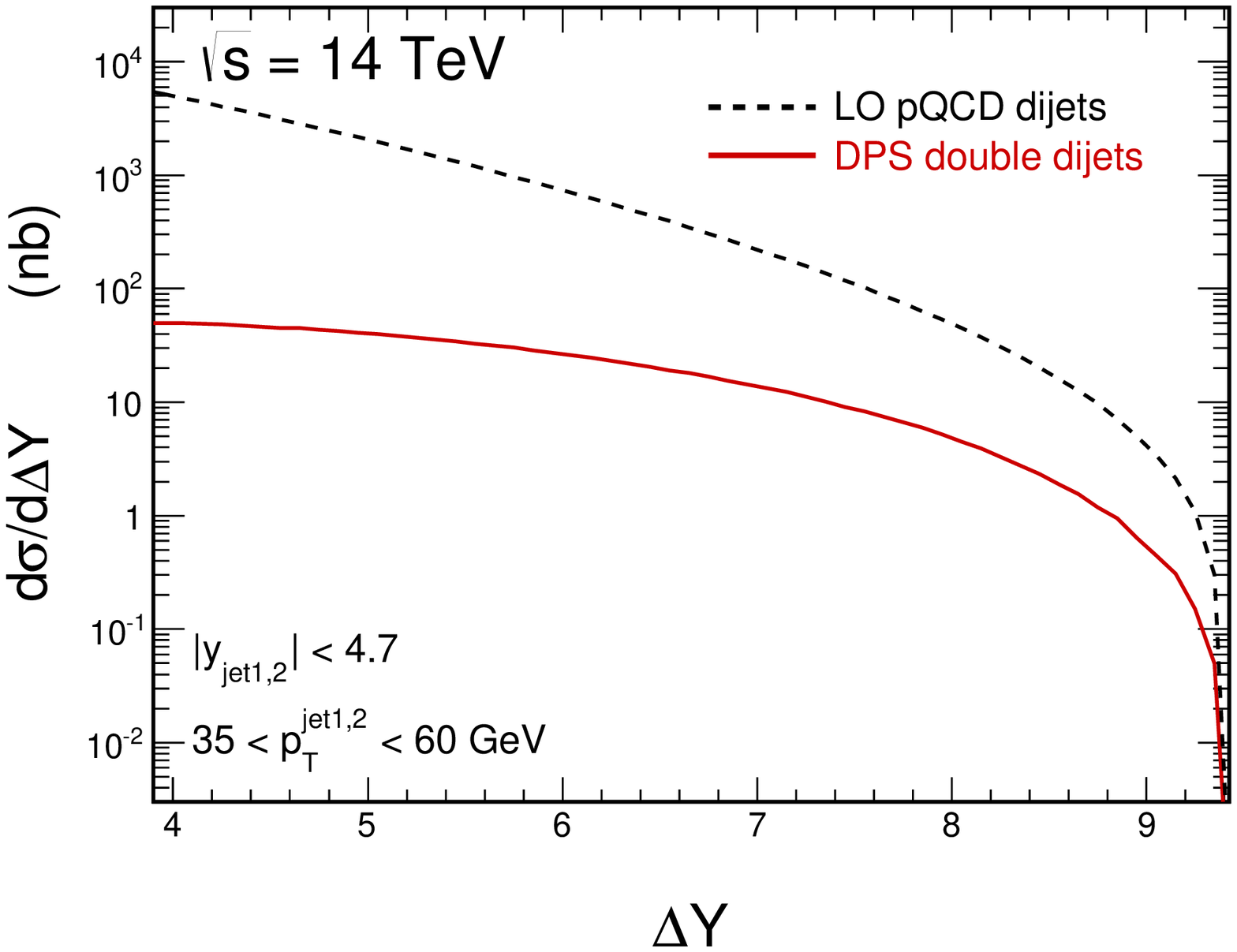}
\end{center}
\vskip -0.3cm
   \caption{
\small Distribution in rapidity distance between jets 
(35 GeV $< p_t <$ 60 GeV).
The collinear pQCD result is shown by the short-dashed line
and the DPS result by the solid line for $\sqrt{s}$ = 7 TeV (left panel)
and $\sqrt{s}$ = 14 TeV (right panel). For comparison we show also
results for the BFKL Mueller-Navelet jets in leading-logarithm 
and next-to-leading-order logarithm approaches from 
Ref.~\cite{Ducloue:2013hia}.
}
 \label{fig:Deltay1}
\end{figure}

In Fig.~\ref{fig:Deltay1} we show distribution in the rapidity 
distance between two jets in LO collinear calculation
and between the most distant jets in rapidity in the case of four DPS jets.
In this calculation we have included cuts for the
CMS expriment \cite{CMS_private}: $y_1, y_2 \in$ (-4.7,4.7),
$p_{1t}, p_{2t} \in$ (35 GeV, 60 GeV).
For comparison we show also results for the BFKL calculation from
Ref.~\cite{Ducloue:2013hia}. For this kinematics the DPS jets
give relatively sizeable contribution only at large rapidity distance.
The NLL BFKL cross section (long-dashed line) is smaller than that for 
the LO collinear approach (short-dashed line).

\begin{figure}[!h]
\begin{center}
\includegraphics[width=5cm]{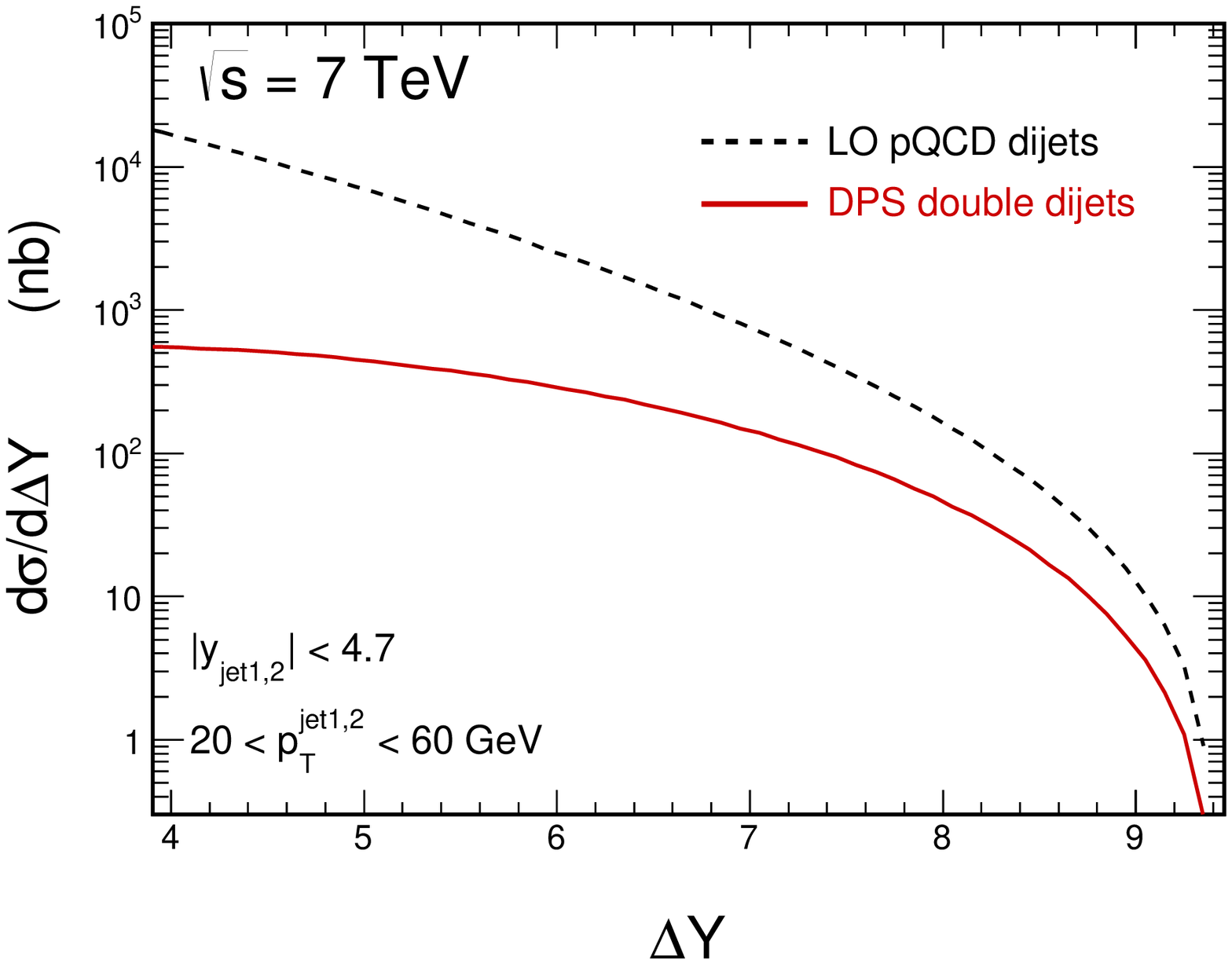}
\includegraphics[width=5cm]{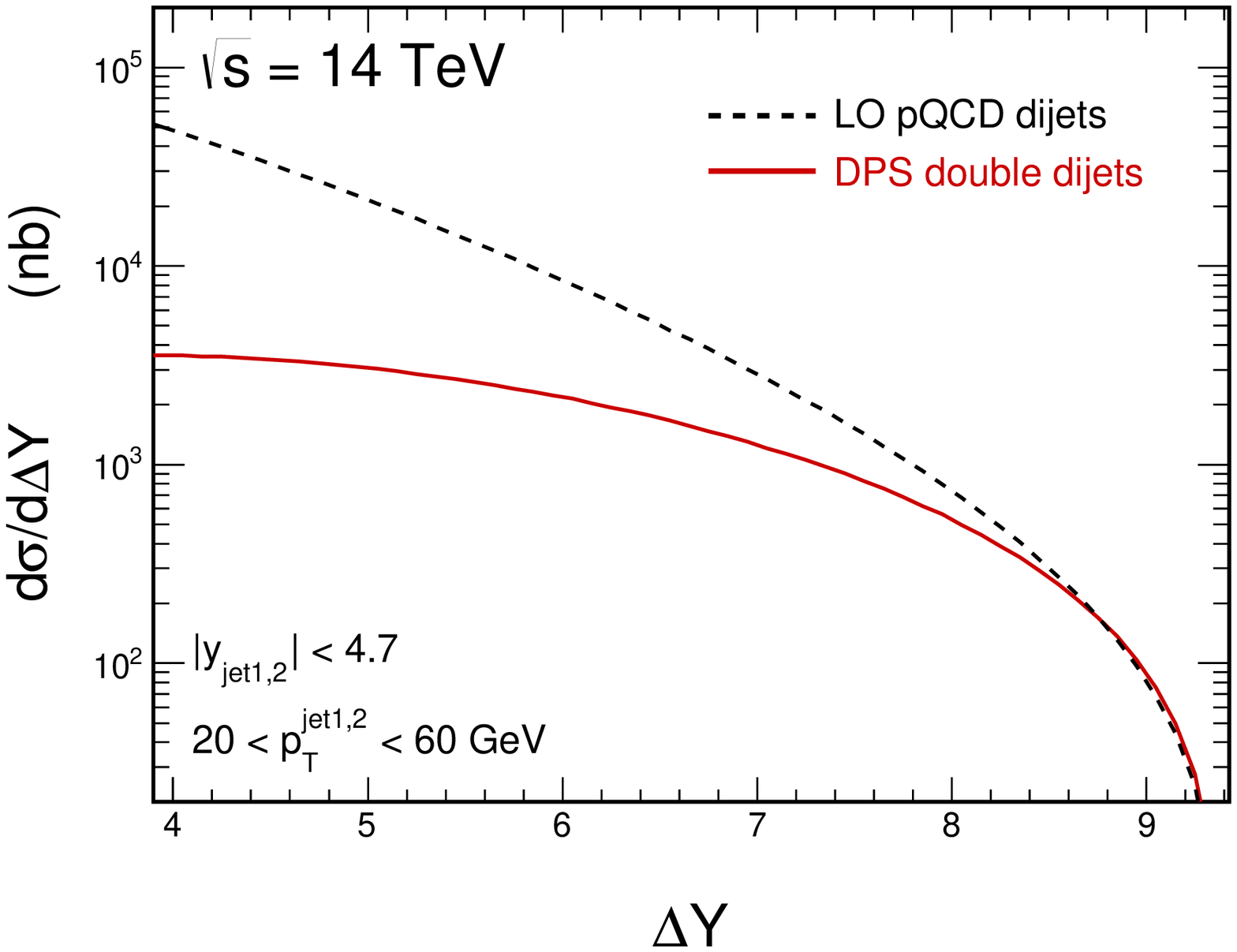} \\
\end{center}
\vskip -0.3cm
\caption{
\small The same as in the previous figure but now for smaller 
lower cut on jet transverse momentum.
}
\label{fig:Deltay-2}
\end{figure}

In Fig.~\ref{fig:Deltay-2} we show rapidity-difference
distribution for even smaller lowest transverse momenta of 
the jet. A measurement of such jets may be, however, difficult. 
Now the DPS contribution may even exceed the standard SPS 
dijet contribution, especially at the nominal LHC energy. 
One could also try to measure correlations of 
semihard ($p_t \sim$ 10 GeV) neutral pions with the help of 
so-called zero-degree calorimeters (ZDC) \cite{MS2014}.

\section{In search for optimal conditions for DPS
contribution in four-jet sample}

\begin{figure}[!h]
\begin{minipage}{0.47\textwidth}
 \centerline{\includegraphics[width=1.0\textwidth]{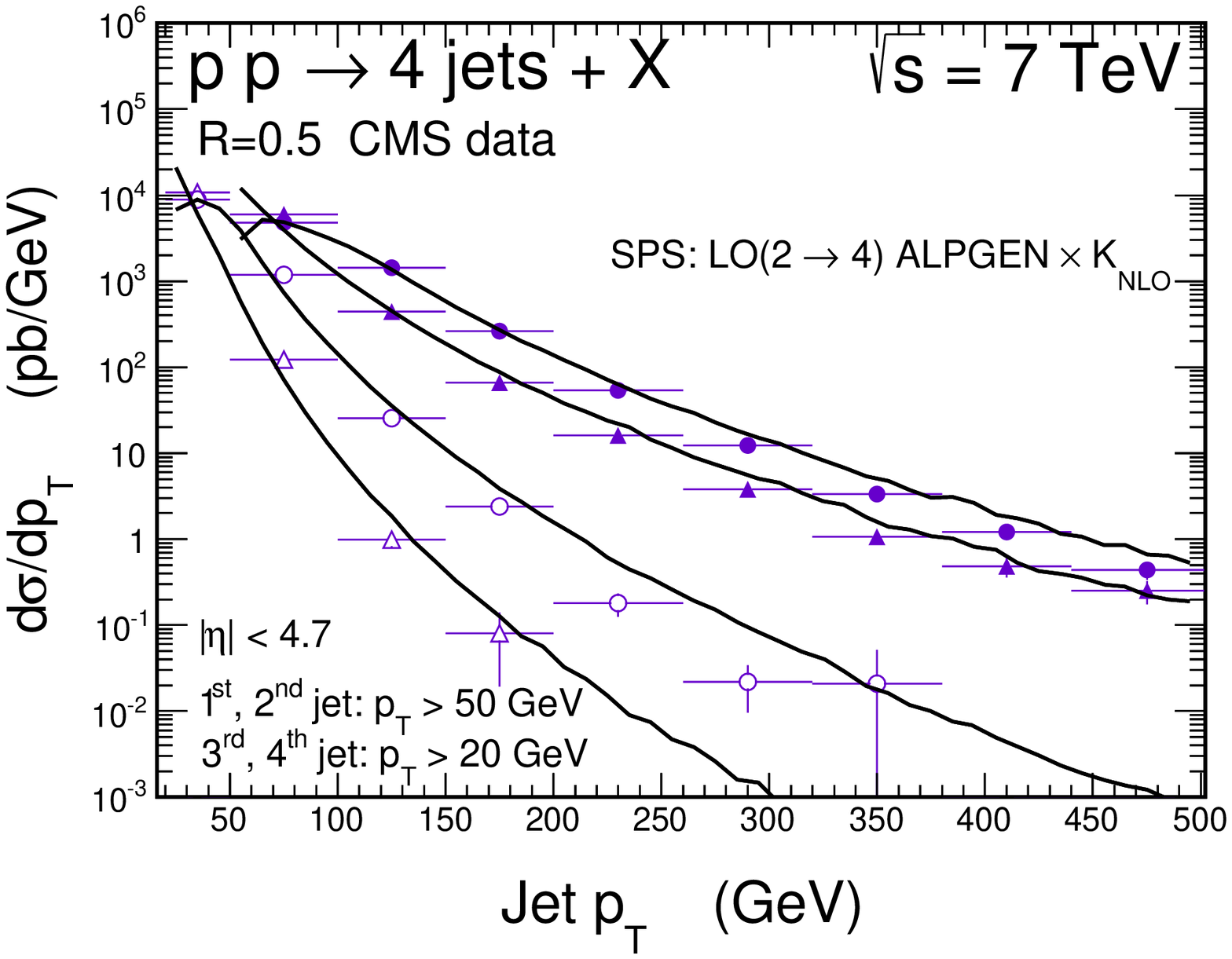}}
\end{minipage}
\hspace{0.5cm}
\begin{minipage}{0.47\textwidth}
 \centerline{\includegraphics[width=1.0\textwidth]{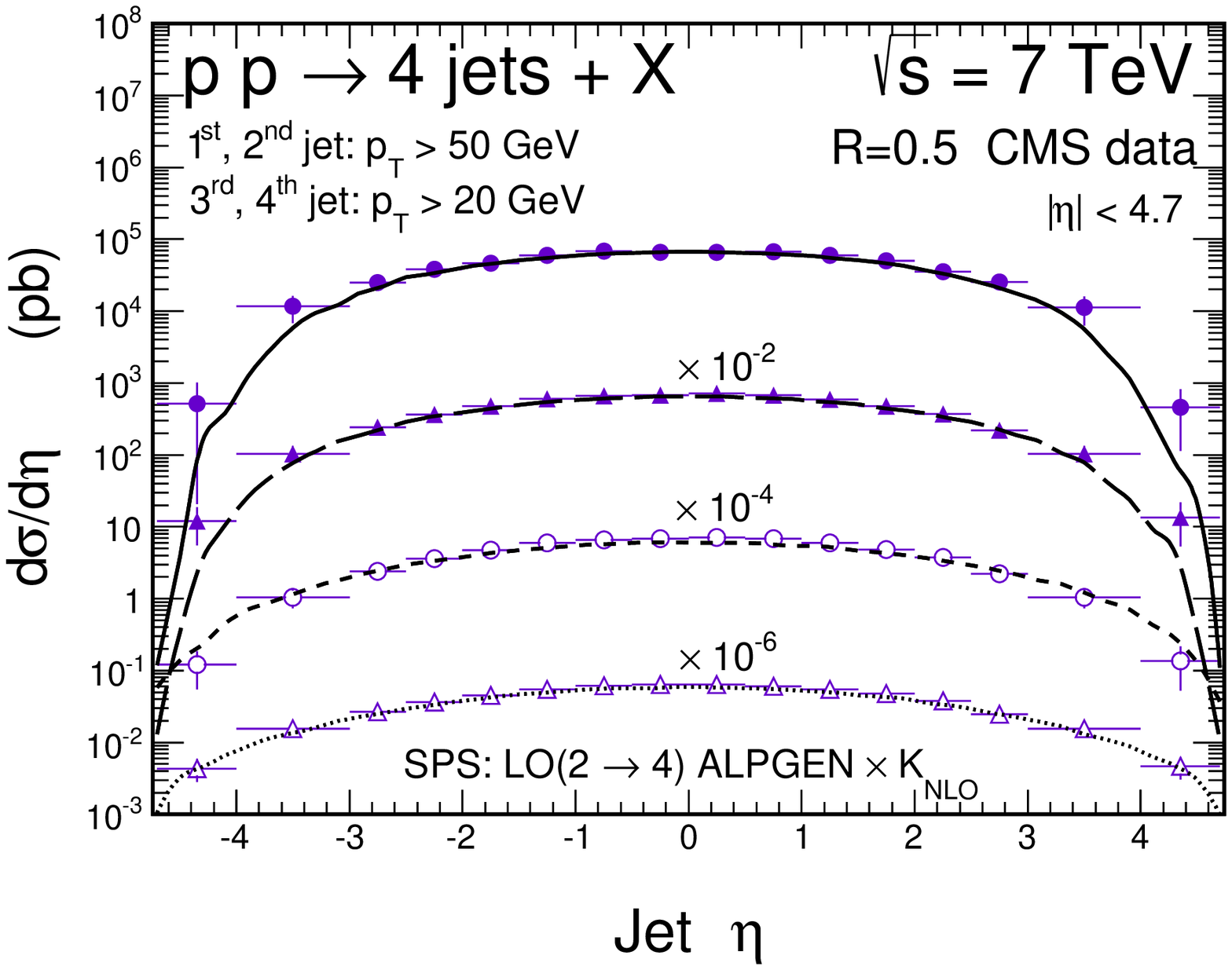}}
\end{minipage}
   \caption{
\small Transverse momentum (left panel) and rapidity (right panel) 
distributions of each of the four-jets (ordered in transverse momentum) 
in the four-jet sample together with the CMS experimental data 
\cite{Chatrchyan:2013qza}. The calculations were performed with
the code ALPGEN \cite{ALPGEN}. Here kinematical cuts relevant for
the experiment were applied to allow for a comparison.}
 \label{fig:CMS-data-1}
\end{figure}

First we wish to demonstrate how reliable our SPS 
four jet calculation is. In Fig.~\ref{fig:CMS-data-1} we compare
the results of calculation with the leading-order code ALPGEN \cite{ALPGEN} 
with recent CMS experimental data \cite{Chatrchyan:2013qza}.
In this analysis the CMS collaboration imposed different transverse momentum
cuts on the leading, subleading, 3$^{rd}$ and 4$^{th}$ jets. 
In this calculation we have used an extra $K$-factor to effectively 
include higher-order effects \cite{Bern:2011ep}.
We get relatively good description of both
transverse moentum and pseudorapidity distributions
of each of the four (ordered in transverse momentum) jets.
Therefore we conclude that the calculation with the ALPGEN generator 
can be a reliable SPS reference point for an exploration of the DPS effects.

\begin{figure}[!h]
\begin{minipage}{0.47\textwidth}
 \centerline{\includegraphics[width=1.0\textwidth]{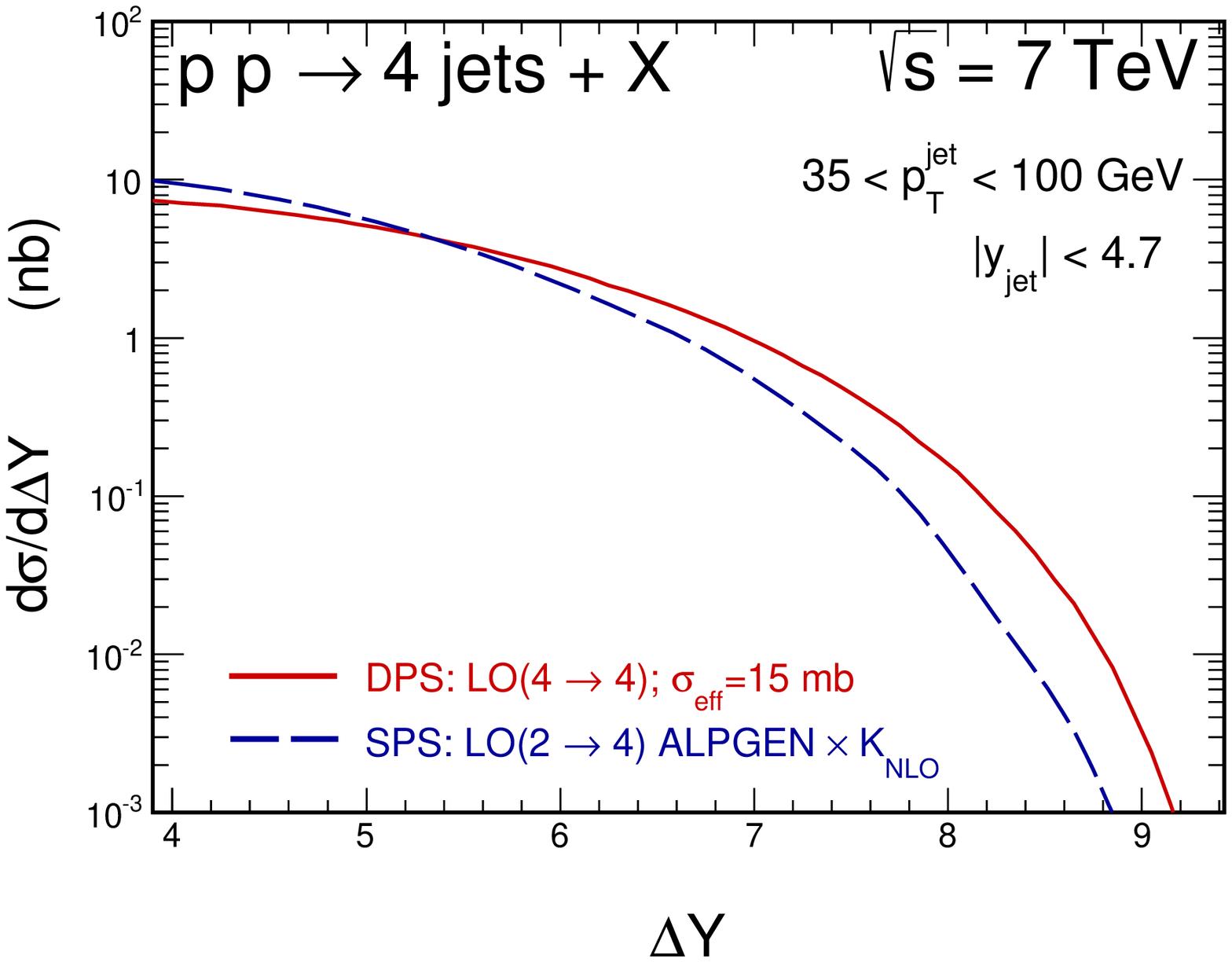}}
\end{minipage}
\hspace{0.5cm}
\begin{minipage}{0.47\textwidth}
 \centerline{\includegraphics[width=1.0\textwidth]{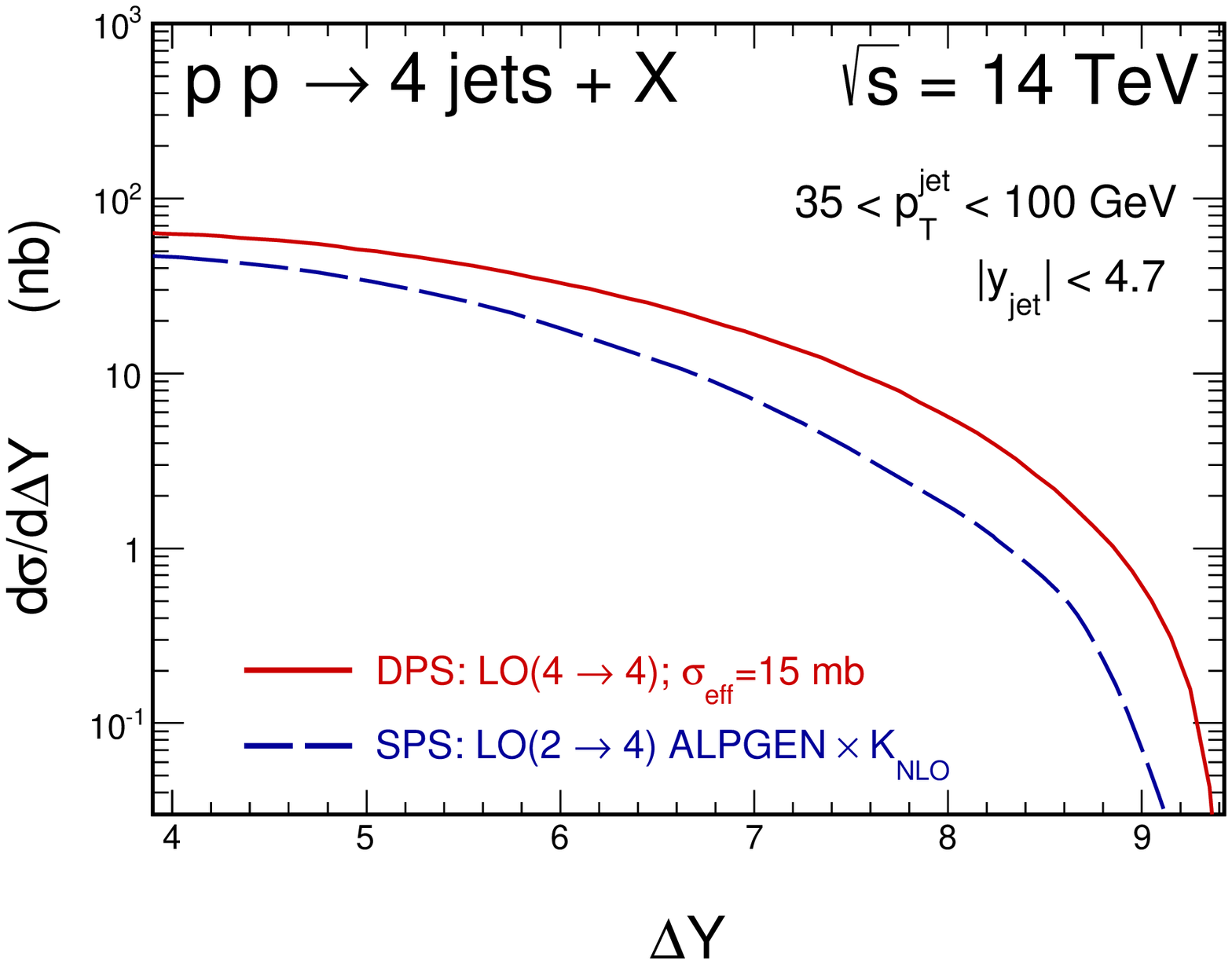}}
\end{minipage}
\begin{minipage}{0.47\textwidth}
 \centerline{\includegraphics[width=1.0\textwidth]{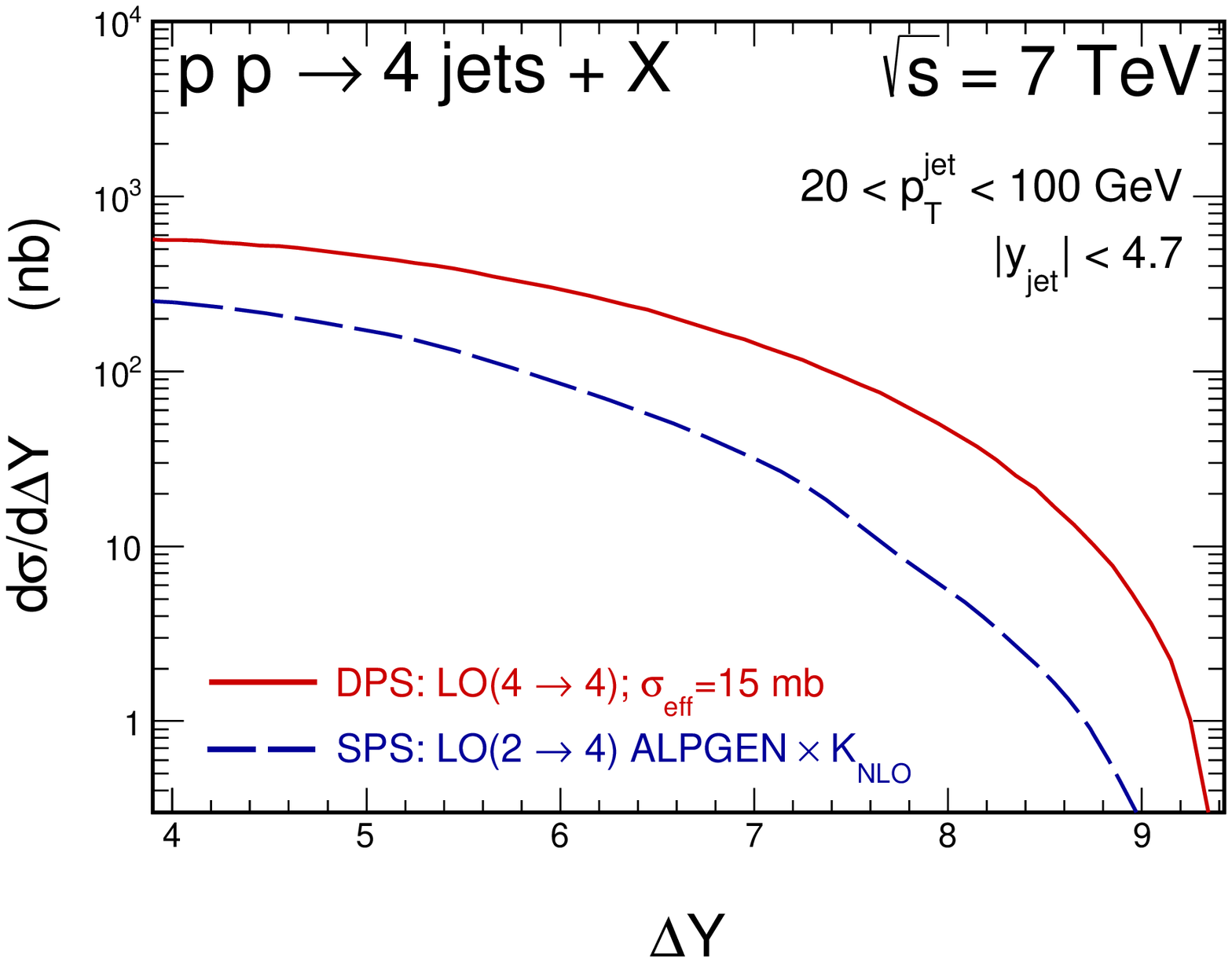}}
\end{minipage}
\hspace{0.5cm}
\begin{minipage}{0.47\textwidth}
 \centerline{\includegraphics[width=1.0\textwidth]{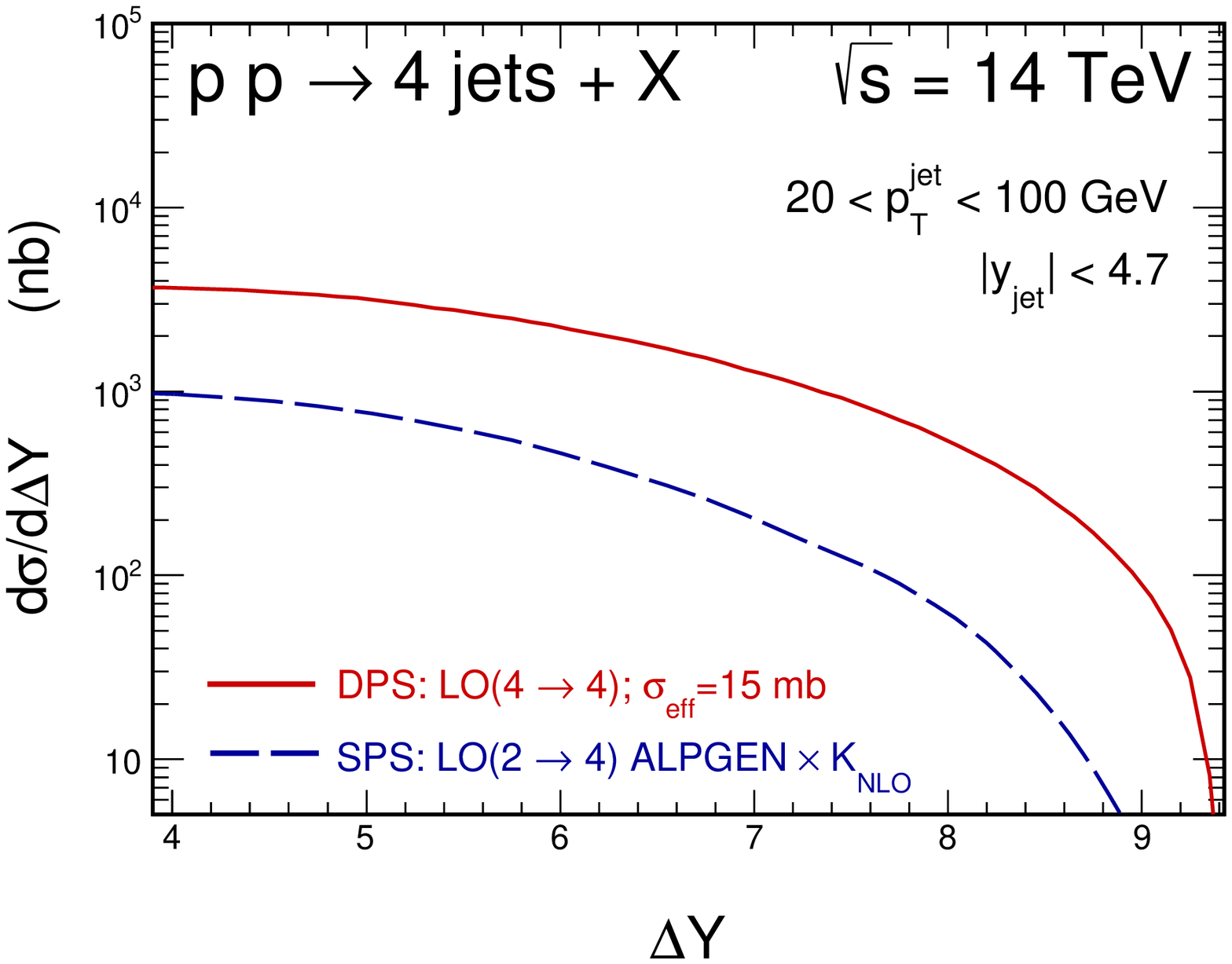}}
\end{minipage}
   \caption{
\small Distribution in rapidity distance of the most remote jets for 
the four-jet sample for $\sqrt{s}$ = 7 TeV (left column) and $\sqrt{s}$
= 14 TeV (right column) for different cuts on jet transverse momenta
(identical for all four jets). }
 \label{fig:DeltaY-DPS-1}
\end{figure}

Having shown that our approach is consistent with existing LHC four-jet data
we wish to discuss how to find optimal conditions for ``observing'' 
the DPS effects. As shown in our previous paper \cite{MS2014}
on dijets widely separated in rapidity the distribution 
in rapidity separation of such jets seems a very good observable 
for observing the DPS.
In Fig.~\ref{fig:DeltaY-DPS-1} we show some
examples of such distributions for different cuts on the jet transverse
momenta for two collision energies $\sqrt{s}$ = 7 TeV and 
$\sqrt{s}$ = 14 TeV obtained
with the condition of the four-jet observation. We focus only
on the distance between the most remote jets and do not check what
happens in between. 
The higher collision energy or the smaller the lower
transverse momentum cut the bigger the relative DPS contribution.
Here the relative DPS contribution is much bigger than for jets
widely separated in rapidity (compare with Fig. \ref{fig:Deltay1} and 
\ref{fig:Deltay-2}).
In such a case one can therefore expect a considerable deficit when only
SPS four jets are included. Such cases would be therefore useful
to "extract" the $\sigma_{eff}$ parameter.
Any deviation from the "canonical" value of 15 mb
would therefore shed new light on the underlying dynamics.
For example, a two-component model discussed in
Refs.~\cite{Gaunt2012,Gaunt:2014rua} strongly suggests such dependences.

\section{Conclusion}

We have discussed how the double-parton scattering
effects may contribute to large-rapidity-distance dijet correlations.
As an example we have shown distributions in rapidity distance between
the most-distant jets in rapidity. The relative contribution of 
the DPS mechanism increases with increasing distance 
in rapidity between jets.
We have also shown some predictions of the Mueller-Navelet jets
in the LL and NLL BFKL framework.
For the CMS configuration our DPS contribution is smaller than 
the dijet SPS contribution, but
only slightly smaller than that for the NLL BFKL calculation.
We have shown that the relative effect of DPS can be increased
by lowering the jet transverse momenta.

In this presentation we have also discussed how to enhance the relative
contribution of double-parton scattering for four-jet production.
First we have confronted results of our calculations with those 
obtained at the LHC by the CMS collaborations.
The comparison indicates some evidence 
of DPS at large pseudorapidities of the leading jet.

We have shown that imposing a lower cut on transverse momenta and 
rapidity distance between the most remote jets improves 
the situation considerably enhancing the relative contribution of DPS.
A dedicated analysis of the DPS effect is possible already with 
the existing data sample at $\sqrt{s}$ = 7 TeV. The situation 
at larger energies, relevant for LHC Run 2 should be even better.
As a consequence we predict that azimuthal correlation between jets 
widely separated in rapidity should dissapear in the considered 
kinematical domain \cite{MS2015}.
We have found that in some corners of the phase space the DPS
contribution can go even above 80 \% \cite{MS2015}.

In this presentation we have presented the detailed predictions. 
Once such cross sections are measured, one could try to extract 
the $\sigma_{eff}$ parameter from the four-jet sample
and try to obtain its dependence on kinematical variables.
Such dependence can be expected due to several reasons such as 
parton-parton correlations, hot spots, perturbative parton splitting.


\end{document}